\documentclass[aps,prb,twocolumn,superscriptaddress]{revtex4-2}

\usepackage{epsfig}
\usepackage{color}
\usepackage{amsmath}
\usepackage{amsfonts}
\usepackage{natbib}
\usepackage{notes2bib}
\usepackage{booktabs}
\usepackage{multirow}
\usepackage{placeins}

\begin{document}

\title{Reciprocal Space Attention for Learning Long-Range Interactions}

\author{Hariharan Ramasubramanian}
\affiliation{Department of Mechanical Engineering, Carnegie Mellon University, Pittsburgh, Pennsylvania 15213}

\author{Alvaro Vazquez-Mayagoitia}
\affiliation{%
Computational Science Division, Argonne National Laboratory, Lemont, IL 60439
}%

\author{Ganesh Sivaraman}
\email[]{ganeshsi@buffalo.edu}
\altaffiliation{Corresponding author}
\affiliation{ 
Department of Materials Design and Innovation, University at Buffalo, Buffalo, NY 14260 \\
}%

\author{Atul C. Thakur}
\email[]{atthakur@ucsd.edu}
\altaffiliation{Corresponding author}
\affiliation{Aiiso Yufeng Li Family Department of Chemical and Nano Engineering, University of California, San Diego, CA 92093}

\begin{abstract}
Machine learning interatomic potentials (MLIPs) have revolutionized the modeling of materials and molecules by directly fitting to \emph{ab initio} data. However, while these models excel at capturing local and semi-local interactions, they often prove insufficient when an explicit and efficient treatment of long-range interactions is required. To address this limitation, we introduce Reciprocal-Space Attention (RSA), a framework designed to capture long-range interactions in the Fourier domain. RSA can be integrated with any existing local or semi-local MLIP framework. 
The central contribution of this work is the mapping of a linear-scaling attention mechanism into Fourier space, enabling the explicit modeling of long-range interactions such as electrostatics and dispersion without relying on predefined charges or other empirical assumptions. We demonstrate the effectiveness of our method as a long-range correction to the MACE backbone across diverse benchmarks, including dimer binding curves, dispersion-dominated layered phosphorene exfoliation, and the molecular dipole density of bulk water. Our results show that RSA consistently captures long-range physics across a broad range of chemical and materials systems. The code and datasets for this work is available at \href{https://github.com/rfhari/reciprocal_space_attention}
{\url{https://github.com/rfhari/reciprocal_space_attention}}.

\end{abstract}

\maketitle

\raggedbottom

\section{Introduction}

Machine learning interatomic potentials (MLIPs) are data-driven surrogates for the potential energy surface trained on \emph{ab initio} energies and forces~\cite{behler2021machine,ceriotti2022beyond}. By directly fitting electronic-structure datasets into high-dimensional models containing on the order of $10^4$--$10^6$ parameters, MLIPs deliver accuracy comparable to density functional theory (DFT) at orders of magnitude lower computational cost per step. This efficiency enables molecular dynamics (MD) simulations of systems with thousands of atoms over nanosecond to microsecond timescales, extending the reach of atomistic modeling to mesoscopic regimes that are otherwise impractical with direct \emph{ab initio} methods~\cite{kovacs2021linear,deringer2021gaussian,lu202186}.

Two families of MLIPs have been especially influential. The first employs physics-inspired SE(3) invariant local descriptors—such as ACSFs, SOAP, and ACE—within kernels or neural networks~\cite{behler2007generalized,bartok2010gaussian,drautz2019atomic}. The second relies on message-passing graph neural networks with explicit rotational equivariance, such as NequIP and MACE, which construct features on the fly while respecting symmetry constraints~\cite{batatia2022mace,batzner20223,batatia2025design}. In both cases, representations are fundamentally local: atoms interact only within a cutoff radius $r_{\mathrm{cut}}$, and for message passing with $L$ layers, the receptive field extends to approximately $L \times r_{\mathrm{cut}}$, providing what is effectively a ``semi-local'' coverage. When trained on large and heterogeneous \emph{ab initio} datasets, these models have demonstrated strong transferability across different chemistry and phases~\cite{batatia2024foundationmodelatomisticmaterials,levine2025open}.

Local and semi-local MLIPs work remarkably well in homogeneous bulk systems such as crystals and simple liquids~\cite{loew2025universal, kaplan2025foundational, du2025universal}, where slowly varying long-range contributions either cancel by symmetry or can be treated in a mean-field sense~\cite{rodgers2008local, rodgers2011efficient, gao2023local, cox2020dielectric}. However, many application-relevant chemical settings—surfaces, interfaces, nanostructures, molecular adsorption, charged or polar media, and systems under external fields—are dominated by intrinsically nonlocal physics, including Coulombic interactions, polarization/induction, and dispersion~\cite{Anstine2023, dhattarwal2023dielectric}. The nearsightedness of electronic matter (NEM) clarifies this distinction: the electronic density at a point is determined primarily by the \emph{effective} potential in its vicinity, which itself contains contributions from long-range electric fields~\cite{Gao2022}. Thus, nearsightedness does not imply locality with respect to atomic coordinates. Incorporating explicit long-range interactions into MLIPs is therefore essential for the faithful modeling of heterogeneous environments.

Two broad strategies have been developed to move beyond the strict locality assumptions of MLIPs~\cite{Anstine2023}. The first relies on charge-augmented schemes, as in PhysNet and its successors~\cite{Unke2019, Unke2021, Cheng2025, Ko2021, rumiantsev2025}, which predict atomic charges (or charge-like) observables from local neighborhoods and then compute electrostatics using Ewald or PME-type solvers~\cite{Deserno1998}. While straightforward to implement, these approaches face well-known challenges: atomic ``charges'' are not observables in \emph{ab initio} theory; different density partitioning schemes yield inconsistent labels; locally predicted charges cannot capture physics beyond the cutoff (e.g., long-range charge transfer); and charge-equilibration fixes introduce additional, often ad hoc, parameters such as electronegativities~\cite{Ko2021}. A second line of work incorporates global interactions directly through fully long-range ML modules. Examples include long-range descriptors (e.g., LODE~\cite{Grisafi2019}, reciprocal-space neural operators (e.g., Neural-P3M, Ewald message passing)~\cite{wang2024neural,kosmala2023}, attention mechanisms with efficient global reach (e.g., SpookyNet, Euclidean Fast Attention)~\cite{frank2024euclidean, Unke2021}, and self-consistent field neural networks (SCFNN)~\cite{Gao2022}.  

From these developments, several key design considerations emerge for incorporating long-range physics into MLIPs: (i) end-to-end differentiability with strict energy–force consistency; (ii) a global receptive field with favorable scaling; (iii) natural compatibility with periodic boundary conditions; (iv) seamless integration with short-range MLIP backbones; and (v) avoidance of non-observables such as predefined atomic charges.

In this work, we introduce Recriprocal Space Attention (RSA), a purely data-driven long-range framework that maps a linear-scaling attention mechanism into Fourier space, providing a global interaction channel while preserving end-to-end differentiability. Our RSA kernels integrate seamlessly with existing short-range MLIPs; in this study we pair it with MACE to construct a unified energy model. By operating directly in the Fourier domain, the module captures electrostatics and dispersion without relying on empirical charge partitioning, while remaining naturally compatible with periodic boundary conditions. We evaluate the approach on benchmarks designed to probe diverse types of long-range physics - including S$_{\rm N}$2 reaction system, dimer binding curves, gas of random charges, liquid NaCl, dispersion-controlled exfoliation in layered phosphorene, and molecular dynamics of liquid water — demonstrating systematic improvements over local and semi-local baselines in energies, forces, and physically relevant observables.

\section{Theory}

\subsection{Real Space Attention}

\label{sec:theory_of_fourier_attention}
We begin by providing the necessary background for establishing a Reciprocal space–based attention mechanism that enables a computationally efficient and accurate treatment of long-range interactions in short-range MLIPs. In particular, our methodology extends naturally beyond real space and remains applicable to periodic systems.

The standard dot-product self-attention mechanism, widely used in transformers, processes inputs through alternating self-attention and feed-forward blocks, with positional information typically introduced via absolute positional embeddings ~\cite{vaswani2017attention}. In this formulation, let $m$ index an item in a sequence of length $N$. The feature at position $m$ is combined with a positional embedding and linearly projected to a query vector $\mathbf{Q}_m$, while tokens at positions $n$ are projected to key vectors $\mathbf{K}_n$ and value vectors $\mathbf{V}_n$. The attention output at position $m$ is then given by
\begin{equation}
\text{A}_m(\mathbf{Q, K, V}) =
\frac{\sum_{n=1}^N \exp\left( \langle \mathbf{Q}_m, \mathbf{K}_n \rangle \right) \mathbf{V}_n}
{\sum_{n=1}^N \exp\left( \langle \mathbf{Q}_m, \mathbf{K}_n \rangle \right)}
\label{n2_attention}
\end{equation}

where \(\mathbf{Q}\in\mathbb{R}^{N\times d_k}\), \(\mathbf{K}\in\mathbb{R}^{N\times d_k}\), and \(\mathbf{V}\in\mathbb{R}^{N\times d_v}\) are the query, key, and value vectors respectively, with N tokens and $d_k$ features; \(\langle \mathbf{Q}_m,\mathbf{K}_n\rangle=\mathbf{Q}_m^\top \mathbf{K}_n\) denotes the scalar dot product. 

A key limitation of Eq.~\ref{n2_attention} is its quadratic complexity, $\mathcal{O}(N^2)$, both in computation and in memory, as each of the $N$ queries attends to all $N$ key vectors. A further limitation is the reliance on absolute positional encoding which directly incorporates positional representations. 

To address the latter issue, recent work has introduced relative positional encoding, which better reflect the pairwise relationships between tokens. Rotary positional embeddings (RoPE)~\cite{Su2024}, first introduced in \textsc{RoFormer}, implement this idea by applying position-dependent rotations $\mathbf{R}_m$ to the query and key vectors. This approach has been shown to improve performance in tasks where relative structure is essential. In parallel, efforts to reduce the quadratic computational cost have led to the development of linear attention mechanisms, which approximate the softmax kernel using feature maps that enable linearization~\cite{katharopoulos2020transformers}. 

To make RoPE compatible with such linear transformers, approximations have been proposed that combine position-dependent rotations with kernel feature maps, allowing relative positional information to be retained while preserving $\mathcal{O}(N)$ complexity. The RoPE integrated linear attention is then

\begin{equation}
\text{A}_m(\mathbf{Q, K, V}) =
\frac{\sum_{n=1}^N (\mathbf{R}_m \phi(\mathbf{Q}_m))^T \, \mathbf{R}_n \phi(\mathbf{K}_n) \mathbf{V}_n}
{\sum_{n=1}^N \phi(\mathbf{Q}_m)^T \, \phi(\mathbf{K}_n)}
\label{eq:quadratic_attention}
\end{equation}
\\
where $\phi$ denotes a feature map (typically non\mbox{-}negative). For each query vector $\mathbf{Q}_m$, the quantity
$\sum_{n=1}^{N} \phi(\mathbf{K}_n)\,\mathbf{V}_n^{\top}$ is independent of $m$ and can be pre-computed as $\mathbf{K}-\mathbf{V}$ cache. Consequently,
Eq.~\ref{eq:quadratic_attention} can be written as

\begin{equation}
\text{A}_m(\mathbf{Q, K, V}) =
\frac{(\mathbf{R}_m \phi(\mathbf{Q}_m))^T \, \sum_{n=1}^N \, \mathbf{R}_n \phi(\mathbf{K}_n) \mathbf{V}_n^T}
{\phi(\mathbf{Q}_m)^T \sum_{n=1}^N \phi(\mathbf{K}_n)}
\label{eq:linear_attention_v2}
\end{equation}
\\
which makes the overall computation linear via $\mathcal{O}(N)$. We encourage the interested reader to refer to the original papers for more details on linear attention and RoPE~\cite{Su2024,katharopoulos2020transformers}. In subsequent discussions, we focus on integrating Eq.~\ref{eq:linear_attention_v2} into atomic systems subject to periodic boundary conditions.

\subsection{Reciprocal Space Periodic Attention (RSA)}

Building on the real-space attention framework discussed in Sec.~\ref{sec:theory_of_fourier_attention}, we now extend the formulation into reciprocal (Fourier) space, which naturally accommodates periodic boundary conditions and long-range interactions. 
In molecular dynamics (MD) simulations, atoms interact with their neighbors not only through real-space short-range interactions but also via reciprocal-space long-range interactions, such as electrostatics, dipole--dipole couplings, dispersion, etc. Capturing these long-range interactions in a purely real-space setting necessarily incurs $\mathcal{O}(N^2)$ cost in both compute and memory, unless approximations such as multipole expansions are introduced~\cite{Greengard1987, gibbon2002long}. 
In contrast, reciprocal-space methods, such as Ewald summation and its generalizations, treat slowly decaying interactions more efficiently by decomposing the potential into short- and long-range contributions, making them naturally well suited for a Fourier-space attention mechanism.

Following the classical Ewald partitioning, the total potential $V(r)$ can be decomposed into a rapidly varying Gaussian-truncated short-range (SR) and uniformly slowly varying long-range (LR) piece~\cite{gingrich2010simulating}

\begin{align}
V(r) &= v_\text{SR}(r) + v_\text{LR}(r) \notag \\
&= \frac{\text{erfc}\left(\frac{r}{\sqrt{2}\sigma}\right)}{r} 
+ \frac{\text{erf}\left(\frac{r}{\sqrt{2}\sigma}\right)}{r}
\end{align}

where $\text{erf}$ and $\text{erfc}$ denote the error (${\text{erf}(x) = 2/\sqrt{\pi}} \int_0^x e^{-t^2}~\text{d}t $) and complementary error functions ($\text{erfc}(x) = 1 - \text{erf}(x)$), respectively, and $\sigma$ is the screening parameter that defines the length scale separating short- and long-range contributions. We assume that $v_{\text{SR}}(r)$ can be fully represented by any modern MLIP such as MACE, and therefore focus only on representing $v_{\text{LR}}(r)$.
Following Ewald summation, the long-range interaction energy of a charge-neutral system can then be written as 

\begin{align}
E_{\mathrm{LR}}
&= \frac{2\pi}{V}
\sum_{\mathbf{k} \neq \mathbf{0}}
\frac{e^{-k^{2}\sigma^{2} / 2}}{k^{2}}
\sum_{m}^{N} \sum_{n=1}^{N}
\tilde{q}_{m} \tilde{q}_{n} \,
e^{\, i \mathbf{k} \cdot (\mathbf{r}_{m} - \mathbf{r}_{n})} \notag \\
&= \frac{2\pi}{V}
\sum_{\mathbf{k}\neq \mathbf{0}}
\frac{e^{-k^{2}\sigma^{2}/2}}{k^{2}}
\;\big|S(\mathbf{k})\big|^{2}
\label{eq:E_long}
\end{align}

where $V$ indicates the volume of the simulation box, $N$ is the total number of atoms, $\mathbf{k}$ are the reciprocal lattice vectors, and $\tilde{q}_m, r_m$ indicate the atomic charges and positions of the $m$-{th} atom respectively. By excluding the divergent $ \mathbf{k} = 0 $ term, the charge neutrality of the system is implicitly enforced.

We express the Ewald sum in terms of the magnitude of the structure factor, $\lvert S(\mathbf{k})\rvert^{2}$, in Eq.~\ref{eq:E_long}.
This reformulation is not merely cosmetic: for a fixed reciprocal-space grid, it reduces the complexity from $\mathcal{O}(N^{2})$ to $\mathcal{O}(N)$ where N is the total number of atoms. Looking closely at the expression for $E_{\rm LR}$ in Eq.~\ref{eq:E_long}, one can see that the factor $S(\mathbf{k})S(-\mathbf{k})$ is the only term that globally couples all atoms through the exponential, $e^{i\mathbf{k}(\mathbf{r}_m - \mathbf{r}_n)}$, and thereby ensures correct treatment of long-range interactions.
Moreover, because the reciprocal lattice is periodic, translational invariance is guaranteed, allowing us to operate directly on real-space position vectors without explicitly constructing periodicity-aware edge lists.

By integrating RoPE methodology with the Ewald-sum formalism, we develop a reciprocal space attention kernel that inherently satisfies periodic boundary conditions while preserving translational invariance. This is achieved through the encoding of pairwise interaction via Bloch-like phase factors. We refer to this as Fourier Positional Encoding (FPE), which is defined as

\begin{equation}
    \text{FPE}_k(\textbf{x}, \vec{\mathbf{r}}_m) = \textbf{x} \cdot e^{i \vec{\mathbf{k}} \cdot \vec{\mathbf{r}}_m}
\label{eq:fpe}
\end{equation}

where \(m\) denotes the \(m\)-th atom in the simulation cell. A key advantage of using FPE is its phase invariance in the periodic lattice space, which gives the following

\begin{equation}
    e^{i \mathbf{k} \cdot \left[ (\mathbf{r}_m - \mathbf{r}_n) + \mathbf{T} \right]}
= e^{i \mathbf{k} \cdot (\mathbf{r}_m - \mathbf{r}_n)}
\end{equation}

where $\mathbf{T}$ is a lattice translational vector. 
This shows that the phase factor is \emph{invariant} under the choice of a
periodic image $\mathbf{T}$, provided that $\mathbf{k}$ is a reciprocal lattice vector of the simulation cell. As a consequence, quantities built purely from such phase factors, such as the Fourier transformed density, are inherently periodic. 

\begin{figure*}[tb]
  \centering
  \includegraphics[width=0.9\linewidth]
  {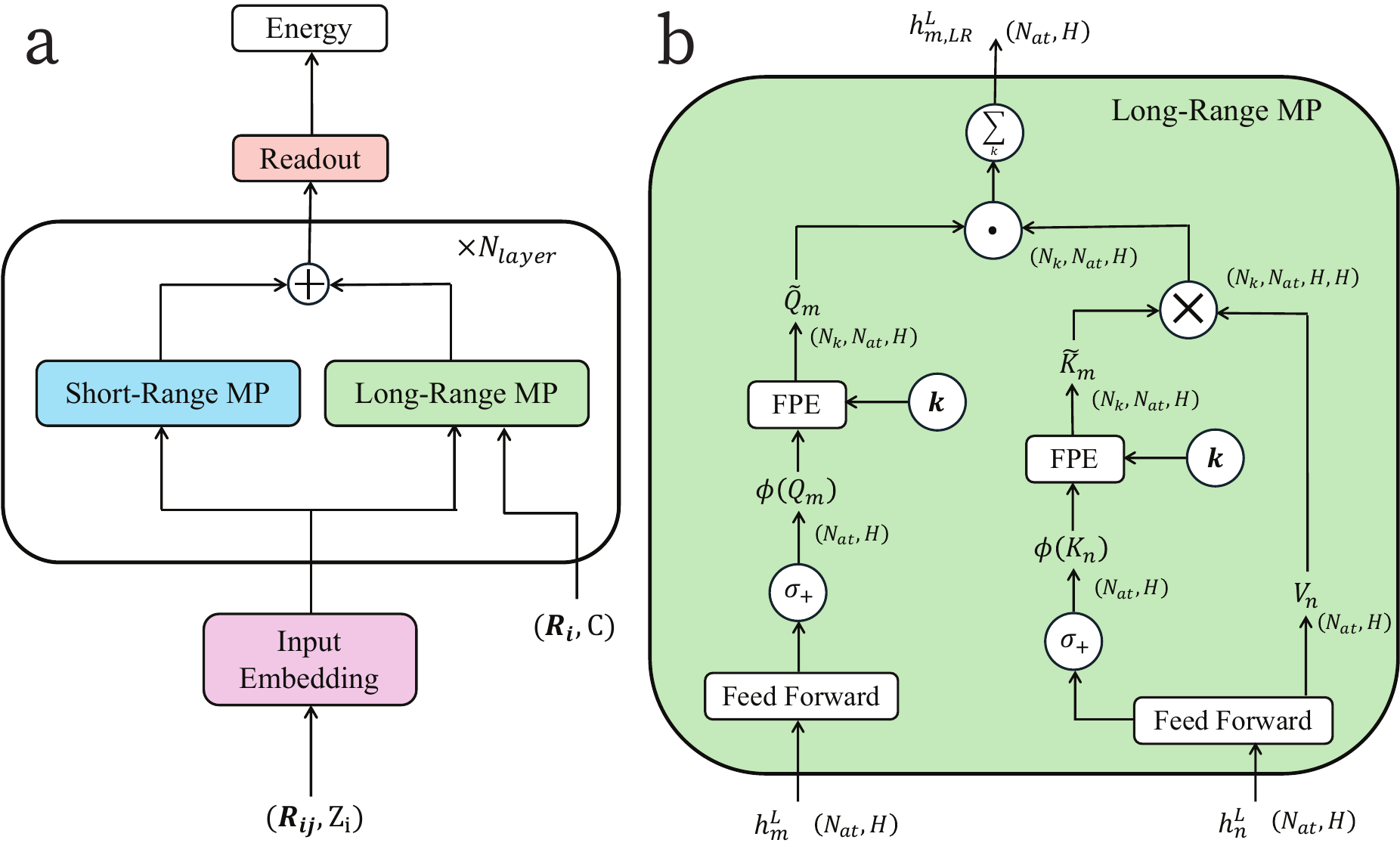}
  \caption{(a) Overview of the short-range and long-range GNN architecture. Here, $Z_i$ denotes the atomic number of atom $i$, $\mathbf{C}$ are the crystal lattice vectors, $\mathbf{R}_i$ is the position of atom $i$, and $\mathbf{R}_{ij}$ is the displacement vector from atom $j$ to atom $i$ (i.e., $\mathbf{R}_{ij}=\mathbf{R}_j-\mathbf{R}_i$). Each interaction layer couples a short-range message-passing (SR-MP, blue) block with a long-range message-passing (LR-MP, green) block, resulting in a complete atomic representation that are passed through a readout head to predict energy and forces. 
  (b) Schematic depiction of Reciprocal Space Attention (RSA) module, which provides the long-range message-passing channel. Atomic features $h_m^L$ and $h_n^L$ are first projected into query, key, and value vectors through feed-forward layers followed by nonlinearities $\sigma_{+}$ to obtain $\phi(Q_m)$ and $\phi(K_n)$. The transformed feature vectors are then rotated using Fourier Positional Encodings (FPE), parameterized by reciprocal lattice vectors $\mathbf{k}$, yielding the rotated $\tilde{\mathbf Q}_m$ and $\tilde{\mathbf K}_n$ vectors. The rotated keys, $\tilde{\mathbf K}_n$, are combined with the value vector $\mathbf V_n$ via an outer product, $\tilde{\mathbf K}_n \otimes \mathbf V_n$, to obtain a graph-level key-value cache. This cache is broadcast back to the node space and contracted (left-multiplied) with the rotated queries $\tilde{\mathbf Q}_m$ at the corresponding $k$-frequency, producing per-mode interactions for each node. Summation over reciprocal modes $\sum_k$ yields the long-range message. This framework provides a linear-scaling, end-to-end differentiable treatment of long-range interactions that naturally accounts for interactions across periodic images of an atom.
  }
  \label{fig:model_architecture}
\end{figure*}

We can formulate an attention kernel (see Fig.~\ref{fig:model_architecture}) using FPE for query and key vectors. 
Since both the query and key vectors are complex, the scalar product is instead defined as $\langle \mathbf{Q}, \mathbf{K}\rangle = \mathbf{Q}^T{\mathbf{\bar{K}}} $, where $\mathbf{\bar{K}}$ indicates a complex conjugate.
The inner product between the query and key vectors associated with atoms $m$ and $n$ in a standard attention format can then be written as

\begin{equation}
\label{qk}
\big\langle \mathbf{Q}_m e^{i\mathbf{k} \cdot \mathbf{r}_m}, 
\mathbf{K}_n e^{i \mathbf{k} \cdot \mathbf{r}_n} \big\rangle
= \langle \mathbf{Q}_m,\ \mathbf{K}_n \rangle e^{i \mathbf{k} \cdot (\mathbf{r}_m - \mathbf{r}_n)}
\end{equation}

Eq.~\ref{qk} is one of the key advantages of FPE which enables us to write a standard quadratic attention operation without normalization and a row-wise softmax as

\begin{equation}
\begin{aligned}
\operatorname{RSA}_m(\mathbf{Q},\mathbf{K},\mathbf{V}) &= \mathbf{V}_m \\
&= \sum_{\substack{k\neq 0\\ n=1,\dots,N}}
   (\mathbf{Q}_m,\mathbf{K}_n)\, e^{i\,\mathbf{k}\cdot(\mathbf{r}_m-\mathbf{r}_n)}\, \mathbf{V}_n
\end{aligned}
\label{eq:vm}
\end{equation}
\\
where we reduce over the $\mathbf{k}$ dimension to get back the final attention outputs $\mathbf{V}_m$. This expression strongly resembles the total long-range potential for an atom $m$ given by Ewald sum~\cite{Loche2025, gingrich2010simulating}

\begin{equation}
    V_{m}^{\rm LR} = 
    \frac{2\pi}{V}
    \sum_{\mathbf{k \neq 0}} \sum_{n=1}^{N} 
    \frac{e^{-k^{2}\sigma^{2} / 2}}{k^{2}}
    \tilde{q}_n e^{i\mathbf{k}(\mathbf{r}_m-\mathbf{r}_n)}
\end{equation}

To ameliorate the cost associated with quadratic attention, we can approximate the above expression using a linear attention-like mechanism discussed in Eq. \ref{eq:linear_attention_v2}, the inner summation within the value matrix computation (Eq. \ref{eq:vm}) remains constant for a given atom \textit{m}. This atom-independent property enables a significant reduction in computational complexity, reducing it from a quadratic to a linear operation. Thus, using a linear attention-like mechanism, the Reciprocal-Space Attention (RSA) formulation can be written as

\begin{align}
    \text{RSA}_m(\mathbf{Q, K, V}) 
    &\simeq \sum_{\mathbf{k \neq 0}} 
    \mathbf{w_k} ~\text{FPE}(\phi(\mathbf{Q}_m), \mathbf{r_m})^T  \notag \\
    &\quad \times \sum_{n=1}^N 
    \text{FPE}(\phi(\mathbf{K}_n), \mathbf{r_n}) \, \mathbf{V_n}^T
\end{align}

where $\mathbf{w}_{\mathbf{k}}=\exp(-k^{2}\sigma^{2}/2)/k^{2}$ are the Ewald–sum based weights for each wavevector $\mathbf{k}$, and $\mathrm{FPE}(\phi(\mathbf{Q}_m),\mathbf{r}_m)$ and $\mathrm{FPE}(\phi(\mathbf{K}_n),\mathbf{r}_n)$ are the feature-mapped, FPE-rotated query and key vectors for atoms $m$ and $n$, respectively.

\section{Results}
Several recent machine learning interatomic potentials~\cite{batatia2024foundationmodelatomisticmaterials, mazitov2025petmadlightweightuniversalinteratomic, Unke2021, wood2025umafamilyuniversalmodels} replace or augment standard message passing updates with self-attention mechanisms over neighboring atoms or edges within local atomic environments. While message-passing neural networks (MPNNs) are effective in capturing semilocal interactions over a few hops, they are not expected to represent true long-range effects as discussed in~\cite{Anstine2023}. In the absence of intermediary nodes connecting distant atoms, no information can propagate. In addition, MPNNs often suffer a loss of expressivity as the number of message-passing steps increases (e.g. over-smoothing), as discussed in~\cite{alon2021bottleneckgraphneuralnetworks,cai2020noteoversmoothinggraphneural}.

\subsection{Disconnected Molecular Graphs with Long-range effects}
Motivated by the above considerations, we first evaluate the proposed long-range (LR) model on benchmarks of disconnected molecular graphs. This setting tests whether the model can capture long-range interactions when molecular clusters lack explicit graph connectivity.

\subsubsection{$\text{S}_\text{N}2$ Reaction Systems}

\begin{figure}[tb]
  \centering
  \includegraphics[width=\linewidth]
  {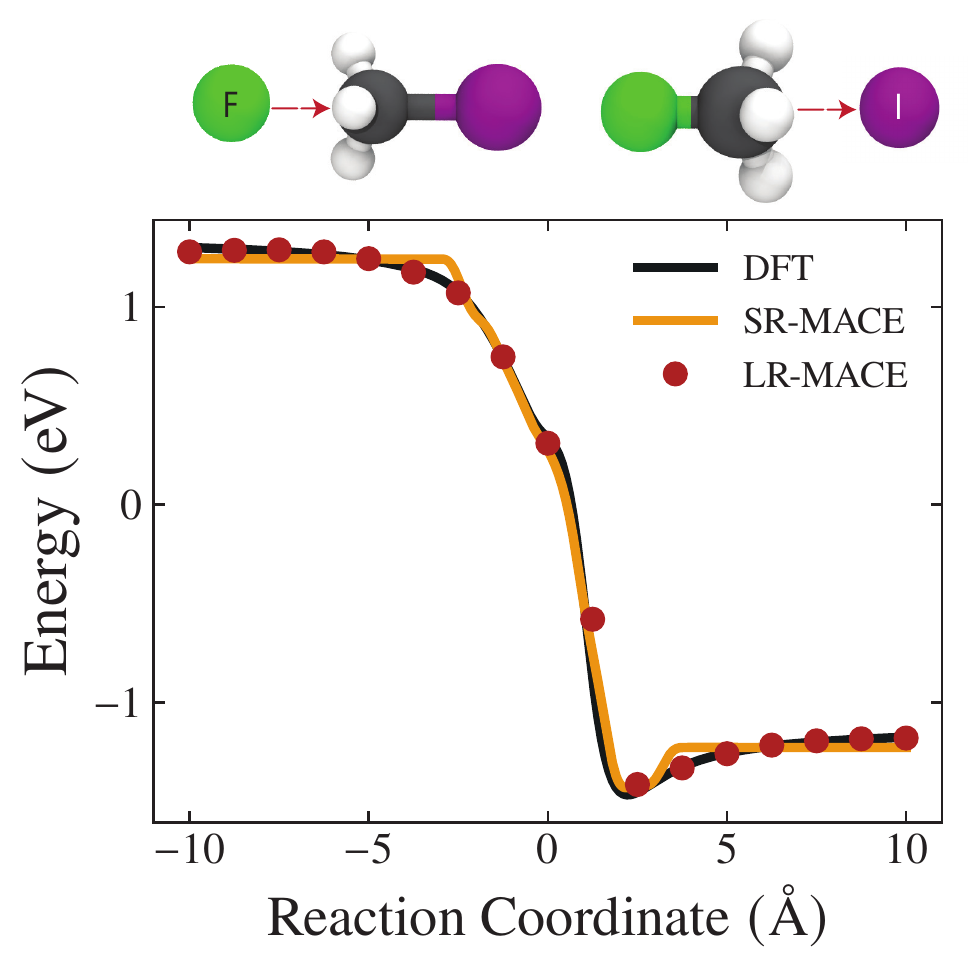}
  \caption{Binding energy curves for the S$_\mathrm{N}$2 reaction complex comparing DFT reference data with predictions from the LR-MACE and SR-MACE models. The LR-MACE model closely reproduces the DFT energy profile along the reaction coordinate, capturing the correct long-range interaction behavior, whereas the SR-MACE model saturates beyond the cutoff due to the absence of explicit long-range interactions.}
  \label{fig:sn2_pes}
\end{figure}

The bimolecular nucleophilic substitution (\(\text{S}_{\text{N}}2\)) is a canonical reaction mechanism in organic chemistry. In this process, a strong nucleophile attacks an \(\mathrm{sp^3}\)-hybridized carbon from the backside, forming a new bond as the leaving group departs in a single, concerted step. As a representative case, we consider reactions involving fluoride and iodide, which are prototypical \(\text{S}_{\text{N}}2\) systems. These systems exhibit pronounced long-range electrostatic interactions between the methyl halide (with a large dipole moment) and the halide anion (bearing a negative charge), posing a significant challenge for strictly local models. We sampled the subset comprising F and I ions from the original \(\text{S}_{\text{N}}2\) data set constructed by~\cite{Unke2019}.

We trained a short-ranged (SR) model using MACE~\cite{batatia2022mace} with a radial cut-off of 5 \AA~and two message passing layers, leading to the total receptive field of 10~\AA. In comparison, a more traditional descriptor-based model is based on local atomic environments that typically use a cutoff of around 5 or 6 \AA, making them even more short-ranged. 

We use the same architectural settings for the long-range (LR) model - two message passing layers with a 5~\AA~cutoff - and with a smearing width ($\sigma = $ 5~\AA).

Fig.~\ref{fig:sn2_pes} presents the one-dimensional potential energy surface along the reaction coordinate. The short-range model exhibits incorrect asymptotic behavior and unphysical artifacts at large ion–molecule separations. As the separation exceeds the cutoff, it predicts a constant energy. In contrast, the LR model accurately captures the potential energy surface across the full range of separations, including the long-distance tail. 

These results show that the LR model remains sensitive to interactions beyond the local receptive field, whereas purely short-range models saturate once the interatomic separations exceed their receptive cutoffs.

\subsubsection{Dimer Binding Curves}
We next benchmark our LR method on binding curves for dimers of charged (C) and polar (P) molecules at varying separations in a periodic cubic box with an edge length of (30~\AA). The dataset is originally derived from the BioFragment Database~\cite{Burns2017}. We additionally select a representative CP dimer class and recompute the curve with PBE0 plus many-body dispersion at finer increments of interatomic distance of 0.1~\AA.

We also computed the potential energy curve of the water dimer (equivalent to the PP dimer case) as a function of the separation distance between the two molecular clusters. These calculations were performed using the SPC/E water model~\cite{Berendsen1987} with explicit long-range Coulomb interactions.

\begin{figure*}[htbp] 
  \centering
  \includegraphics[width=0.9\linewidth]
  {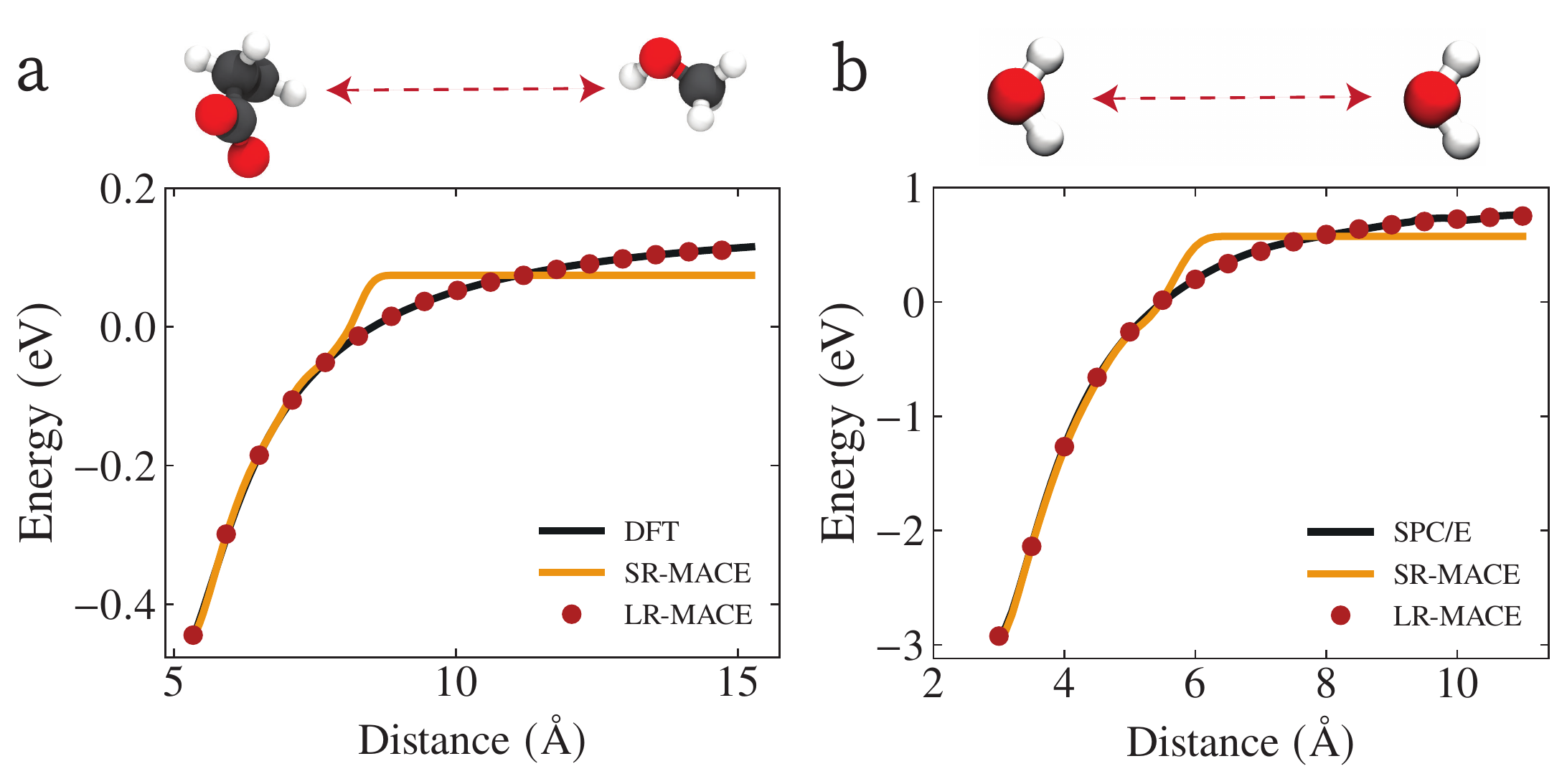}
  \caption{Binding energy curves comparing density functional theory (DFT) reference data with predictions from the LR-MACE and SR-MACE models for (a) the CP dimer and (b) the water dimer. The LR-MACE model closely follows the DFT reference across all intermolecular separations, particularly in the long-range tail of the potential, whereas the SR-MACE model saturates beyond the cutoff range due to the absence of explicit long-range interactions.}
  \label{fig:dimer_binding_curves}
\end{figure*}

Both the SR and LR models use the same architectural settings as in the \(\text{S}_{\text{N}}2\) benchmark. 
Figs.~\ref{fig:dimer_binding_curves}(a) and (b) compare the binding curves for the water and CP dimers, respectively. Consistent with the \(\text{S}_{\text{N}}2\) benchmark, the short-range model saturates once the inter-fragment separation exceeds its total receptive field, whereas the long-range model remains sensitive to interactions beyond this range and recovers the expected long-distance asymptotic behavior.

\subsection{Periodic Graphs with Long-Range effects}
In this section, we evaluate the LR model on established benchmarks that assess its ability to capture long-range interactions under periodic boundary conditions.

\subsubsection{A gas of Random Charges}
As an initial validation, we consider a toy system comprising randomly distributed point charges within a cubic simulation cell with periodic boundary conditions, following the construction by ~\cite{Grisafi2019}. Each configuration contains 128 atoms, of which 64 carry a positive charge of $+1e$ and the remaining 64 carry a negative charge of $-1e$. The atomic interactions are governed by the Coulomb potential, supplemented by the repulsive term of a Lennard-Jones potential. This benchmark is employed to assess the performance of our LR model in comparison to the SR model in an environment with strong electrostatic interactions.

For the SR model, we used a cutoff of 5~\AA~with two message-passing layers. The LR component used the same receptive field and a Gaussian smearing of $\sigma = 5$~\AA. With these settings, the LR model consistently outperformed the SR model, with the improvement most pronounced in the Force MAEs, as shown in Table~\ref{table:random_charge_liquid_nacl}. This indicates that the LR formulation captures per-atom long-range electrostatics more effectively than its short-range counterpart.

\subsubsection{Liquid sodium chloride}
After validating the model on systems governed by $1/r$ interactions, we subsequently assessed its performance on a realistic molten salt system. Liquid sodium chloride (NaCl) was selected as the initial benchmark because of the pronounced electronegativity contrast between Na and Cl atoms, which is expected to induce substantial long-range Coulombic interactions. We employed the dataset of Ref.~\cite{Faller2024}, comprising 1,014 configurations of 128 atoms (64 Na and 64 Cl) each, divided into $80\%$ training and $20\%$ validation splits.

Table~\ref{table:random_charge_liquid_nacl} compares the LR and SR MACE models evaluated with a 6 \AA~receptive field (single message passing layer with $r_\mathrm{cutoff}=6$ \AA). The LR model consistently achieves lower MAEs for both energies and forces, reflecting clear relative reductions and indicating that explicit treatment of long-range interactions, absent in the SR formulation, improves predictive accuracy at the same cutoff. 

\begin{table*}[tb]
\caption{MAEs of Energies and Forces for LR and SR MACE models with corresponding receptive fields.}
\label{table:random_charge_liquid_nacl}
\centering
{
\begin{tabular}{|c|c|c|c|}
\hline
Dataset & Model & Energy MAE (meV/atom) & Force MAE (meV/\AA) \\ \hline
\multirow{2}{*}{Random Charges} & LR MACE (10 \AA) & 2.5 & 71.6 \\ \cline{2-4} 
 & SR MACE (10 \AA) & 3.0 & 97.1 \\ \hline
\multirow{2}{*}{Liquid NaCl} & LR MACE (6 \AA) & 6.8 & 141.9 \\ \cline{2-4} 
 & SR MACE (6 \AA) & 8.7 & 175.1 \\ \hline
\end{tabular}%
}
\end{table*}

\subsubsection{Exfoliation of Phosphorene}

\begin{figure*}[htbp]
    \centering
    \includegraphics[width=0.9\linewidth]
    {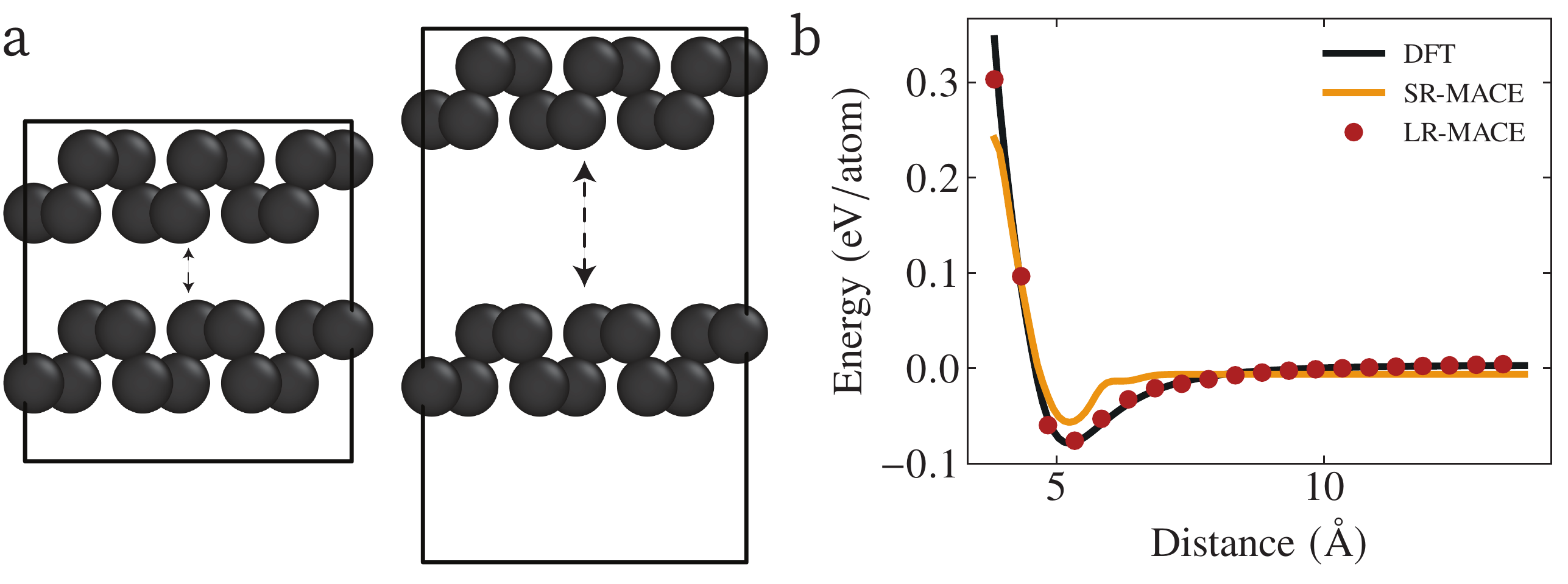}
    \caption{Inter-layer interaction dominated by dispersion. (a) Illustration of Phosphorene sheet exfoliation based on the original work of Deringer \emph{et al.}~\cite{Deringer2020} with the corresponding exfoliation  curve shown in (b). }.
    \label{fig:exfoliate_P}
\end{figure*}

We next investigate the exfoliation  of black phosphorus, focusing on the interaction between pairs of phosphorene layers, building upon the work of~\cite{Deringer2020}. In their study, the exfoliation curve was constructed by starting from the bulk crystal structure of black phosphorus and systematically varying the interlayer distance along the [010] crystallographic direction, while preserving the internal geometry of the puckered monolayers. The interlayer separation was defined as the distance between these layers, and the corresponding energies were computed using density functional theory with many-body dispersion (DFT+MBD). To develop the MLIP, Deringer \emph{et. al.}~\cite{Deringer2020} employed a Gaussian Approximation Potential (GAP) model augmented with an explicit long-range $R^6$ dispersion term. Their analysis revealed that a short-range GAP model lacking the $R^6$ correction failed to reproduce the exfoliation energy profile, especially at large interlayer separations, highlighting the necessity of explicitly including long-range interactions alongside local atomic descriptors.

In this work, we leverage the original dataset from their study to evaluate our long-range machine learning framework. For the short-range (SR) component, we used a cutoff of 6~\AA~with two message-passing layers. The long-range (LR) component employed a Gaussian smearing of $\sigma = 5$~\AA.  The exfoliation energy curve obtained using our method is shown in Fig.~\ref{fig:exfoliate_P}. This plot illustrates the energetic response as the interlayer distance increases, capturing the essential features of van der Waals interactions between phosphorene sheets. Notably, our approach achieves accuracy comparable to DFT+MBD without requiring empirical parameterization of long-range interactions, demonstrating its effectiveness for layered materials.

\subsubsection{Effect of long-range beyond energy and force MAEs — The dynamics of the bulk water}

\begin{figure*}[tb]
    \centering
    \includegraphics[width=0.9\linewidth]
    {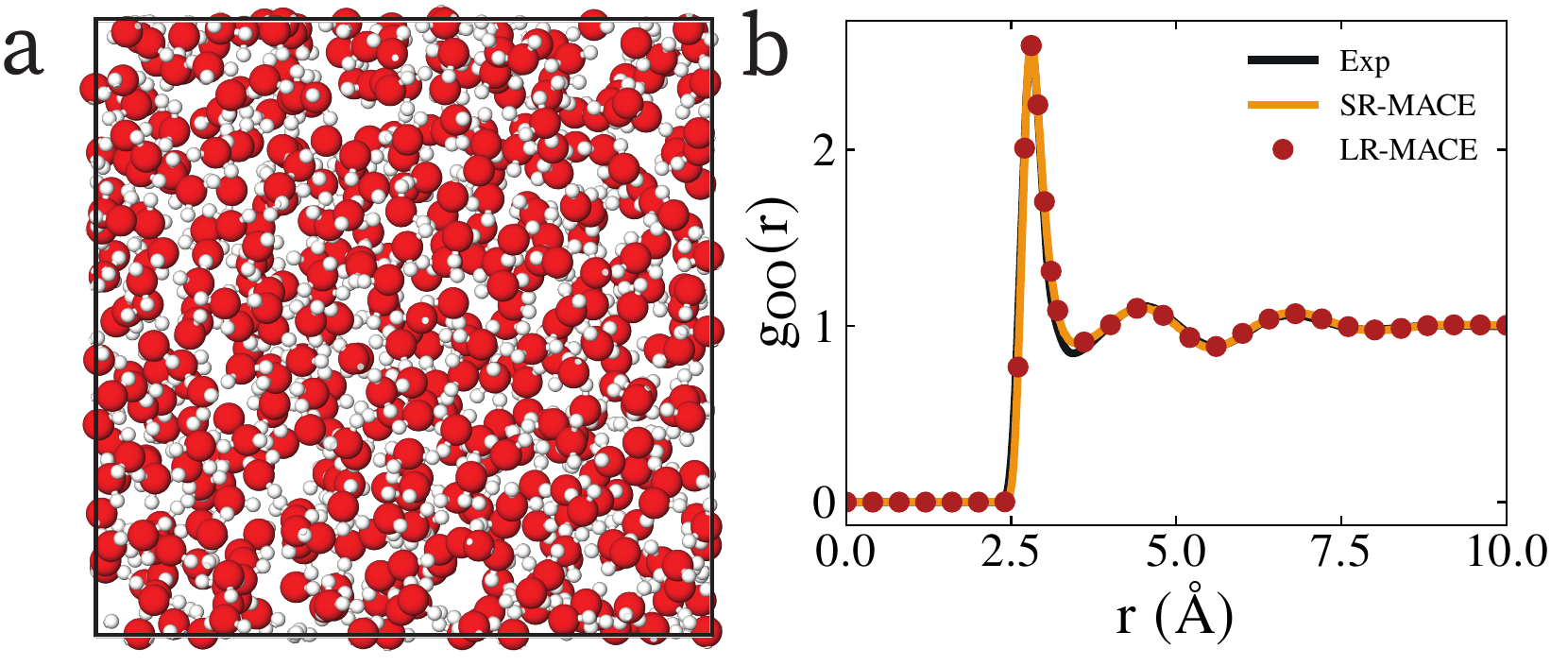}
    \caption{
    (a) Snapshot of bulk water system at 300~K from a molecular dynamics trajectory. (b) Oxygen–oxygen radial distribution function ($\rm g_{\mathrm{OO}}(r)$) comparing the SR-MACE and LR-MACE models. Both models reproduce nearly identical structural features, indicating that the local structure of liquid water is primarily determined by short-range interactions, which the LR model successfully learns directly from the data. The black solid line represents the experimental reference for liquid water~\cite{skinner2014structure}, showing excellent agreement with both models.}
    \label{fig:rdf_oo}
\end{figure*}

\begin{figure}[tb]
    \centering
    \includegraphics[width=\linewidth]
    {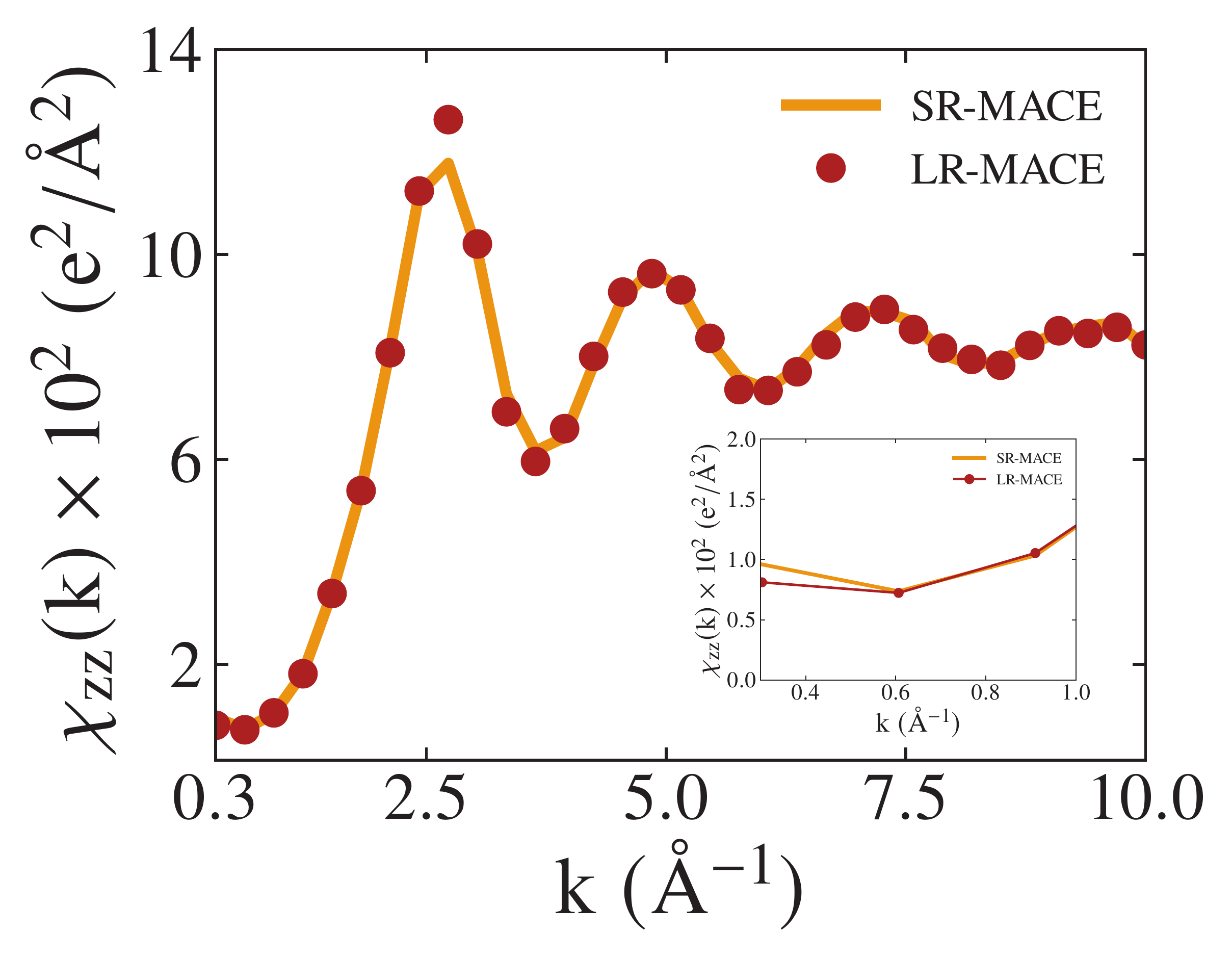} 
    \caption{Longitudinal dipole–density correlation function, $\chi_{zz}(k)$, for bulk water at 300~K obtained from the two-layer short-range (SR-MACE) and long-range (LR-MACE) models. The inset highlights the low-$k$ regime, which is most sensitive to long-range electrostatics. Both models show excellent agreement across all $k$ values except at small $k$, where the LR model correctly reproduces the long-wavelength ($k \to 0$) behavior, while the SR model diverges due to incomplete dipolar screening.}
    \label{fig:dipole}
\end{figure}

To assess the capability of the LR MACE framework in representing complex molecular liquids, we used it to model the structure and dielectric properties of bulk water at 300~K. 
To train both LR and SR models for water, we use a dataset comprising 1593 liquid water configurations ~\cite{Cheng2019}, each containing 64 molecules. The reference configurations in the training data were generated with the CP2K software~\cite{Khne2020} package at the revPBE0-D3 level~\cite{Marsalek2017} of the density functional theory. 
The short-range model uses a cutoff of 6~\AA~with two message-passing layers, which gives it a total receptive field of 12 angstroms. 
The long-range (LR) models is setup identically to the SR model, except that it additionally uses the RSA module with a Gaussian smearing of $\sigma = 5$~\AA.

We used these models to perform MD simulations of bulk water with a density of 1~g/mL at 300 K.
Consistent with previous observations~\cite{Cheng2025, Yue2021}, Fig.~\ref{fig:rdf_oo} shows excellent agreement with the corresponding SR-MACE baseline, suggesting that the LR model can provide stable continuous trajectories crucial for MD simulations.
The agreement between the SR and LR models also highlight that the structure of a uniform bulk liquid is primarily governed by the rapidly varying SR forces and long-range forces have little to no effect on the structure, consistent with the previous observations based on classical SR models~\cite{rodgers2011efficient, rodgers2008localpnas, rodgers2008local, gao2023local, Gao2022}. 
As such, local and semi-local representations can often be adequate for capturing the correct structure of uniform liquids.

Although the overall structure of liquid water is largely insensitive to long-range interactions, the dipolar correlations within the bulk are not~\cite{cox2020dielectric, Cheng2025, gao2023local}. 
The longitudinal component of the dipole–density correlation function, $\chi_{zz}(k)$, which characterizes correlations of the microscopic polarization at wavevector $k$, directly governs the dielectric response. 
In the long-wavelength limit $(k \rightarrow 0)$, $\chi_{zz}(k)$ determines the static dielectric constant of the liquid, making it a stringent test of long-range accuracy.
To examine this, we evaluated $\chi_{zz}(k)$ for both the SR and LR water models. 

As neither the SR nor LR models assign atomic charges explicitly, we evaluated molecular dipoles using the fixed partial charges of the SPC/E water model. 
This choice affects only the overall scale of the dipole moment but not the qualitative behavior of the dipole–density correlations. 
The longitudinal dipole–density correlations, shown in Fig.~\ref{fig:dipole}, exhibit excellent agreement among all three models across most $k$ values, except at long wavelengths (small $k$), where both SR models diverge. 
This divergence indicates incomplete dipolar screening in the SR models, which can lead to significant artifacts when modeling properties such as solvation or dielectric response~\cite{thakur2021distributed}.

We employed a semi-local two-layer model for bulk water, which exhibits a slightly delayed onset of divergence in the dipole–density correlations due to its larger receptive field compared to a fully local, non–message-passing model. 
However, as demonstrated previously~\cite{Cheng2025, Gao2022, cox2020dielectric}, all SR models ultimately fail to capture the correct asymptotic screening behavior.
In contrast, the LR model accurately reproduces the long-wavelength behavior of the dipole–density correlations, consistent with physical expectations for a properly screened polar liquid.

\section{Discussion}
In this work, Reciprocal-Space Attention (RSA) is introduced as a long-range augmentation to semi-local message-passing networks that learns interatomic interactions directly from data, eliminating the need for auxiliary empirical observables (e.g., partial charges or ad hoc dispersion-correction terms). Building on a prototypical RSA formulation, we modify both node embeddings and message updates with lattice-aware relative positional encodings that enumerate all translations of each atom across periodic boundary conditions. 

The RSA framework can be extended in different forms. In this work, the short-range MACE backbone is restricted to scalar channels (\(\ell_{\text{max}}=0\)), although extending to higher-order tensor features is natural. Because the plane-wave factor \(e^{i\mathbf{k}\cdot\mathbf{r}}\) is invariant under simultaneous rotations of \(\mathbf{r}\in\mathbb{R}^3\) and \(\mathbf{k}\in\mathbb{R}^3\) (for any \(R\in\mathrm{SO}(3)\), \(R\mathbf{k}\cdot R\mathbf{r}=\mathbf{k}\cdot\mathbf{r}\)), a rank-\(\ell\) spherical-tensor feature \(T^{(\ell)}_{m}(\mathbf{r})\) can be augmented as
\begin{equation}
\tilde{T}^{(\ell)}_{m}(\mathbf{r},\mathbf{k})
= e^{i\mathbf{k}\cdot\mathbf{r}}\, T^{(\ell)}_{m}(\mathbf{r}).
\label{eq:pfe_extention}
\end{equation}

where \(m\in\{-\ell,\ldots,\ell\}\). In general, equivariance is preserved under the following operations: (i) addition of spherical tensors of the same order \(\ell\); (ii) multiplication by a scalar that is identical across all \(m\) for a given \(\ell\); and (iii) the tensor product of two spherical tensors to form a third via Clebsch-Gordan contraction. In the case of FPE, the exponential factor is an \(m\) independent rotation-invariant scalar, so the irreducible tensor rank \(\ell\) is preserved in Eq.~\ref{eq:pfe_extention} (assuming no inter-\(\ell\) mixing in the RSA calculation). Moreover, RSA is also translation-invariant as it depends on relative displacements, $e^{i\,\mathbf{k}\cdot(\mathbf{r}_m-\mathbf{r}_n)}$, between atom $m$ and atom $n$ as shown in Eq. \ref{eq:vm}.

Although the above construction generalizes to equivariant features, previous work has shown that invariant descriptors are sufficient to capture long-range pairwise interactions~\cite{Loche2025}. Consistent with this, Kosmala \textit{et al.}\ also reported substantial improvements when only the scalar channel was updated within an otherwise equivariant architecture such as PaiNN~\cite{kosmala2023,painn_model}. Guided by these observations, we restrict the present study to a MACE backbone with invariant (scalar) features. A promising extension is a hybrid scheme in which invariant long-range messages are coupled to equivariant short-range messages via tensor products, e.g. \(\mathbf{m}^{\mathrm{LR}}_{\mathrm{inv}} \otimes \mathbf{m}^{\mathrm{SR}}_{\ell m}\), thereby providing global context while preserving the equivariant features to the short-range backbone.

Beyond invariants, the RSA-based long-range message passing method can be generalized to achieve rotational equivariance. This involves using the plane-wave expansion theorem, which states that 
\begin{equation}
e^{i\mathbf{k}\cdot\mathbf{r}}
= 4\pi \sum_{\ell'=0}^{\infty}\sum_{m'=-\ell'}^{\ell'} i^{\ell'}\, j_{\ell'}(kr)\, 
Y_{\ell' m'}(\hat{\mathbf{r}})\, Y^*_{\ell' m'}(\hat{\mathbf{k}}),
\label{eq:plane-wave}
\end{equation}

where $\ell'$ and m' denotes the tensor order, $j_{\ell'}$ is the spherical bessel function of the first kind, $Y^*_{\ell' m'}$ is the complex conjugate of the spherical harmonic, and $\hat{\mathbf{r}},\hat{\mathbf{k}}$ are unit vectors along $\mathbf{r}$ and $\mathbf{k}$, respectively.

Combining Eq.~(13) with Eq.~(12), and for a general form of
$T^{(\ell)}_{m}(\mathbf r) \;=\; f_{\ell}(r)\,Y^{\ell}_{m}(\hat{\mathbf r})$, where $f_{\ell}(r)$ is the radial component, we obtain,
\begin{align*}
\tilde T^{(\ell)}_{m}(\mathbf r,\mathbf k)
&= 4\pi \sum_{\ell'=0}^{\infty}\sum_{m'=-\ell'}^{\ell'} 
   i^{\ell'}\, Y^{*}_{\ell' m'}(\hat{\mathbf k}) \nonumber\\
&\quad \times f_{\ell}(r)\, j_{\ell'}(kr)\,
   Y_{\ell' m'}(\hat{\mathbf r})\, Y_{\ell m}(\hat{\mathbf r}).
\label{eq:tildeT}
\end{align*}

The above expression closely resembles LODE~\cite{Grisafi2019} potential-field descriptors, in which a global field is computed in a reciprocal space to modify the local descriptors. In other words, a key advantage (and similarity) of LODE and RSA is their flexibility, i.e. long-range interactions are not hard parameterized by predefined functional forms, but are represented as long-range field descriptors that couple naturally to local atomic descriptors to get per-atom energies in periodic systems. 

Another avenue for extending the framework is to incorporate mesh-based Ewald solvers, e.g., Particle Mesh Ewald (PME), smooth PME (SPME), or Particle–Particle–Particle–Mesh (PPPM)~\cite{Darden1993, Essmann1995, hockney2021computer, Deserno1998}. In the present formulation, although the reciprocal space attention scales linearly with the number of $\mathbf{k}$-vectors, summation over the full reciprocal lattice reproduces the classical $O(N^{3/2})$ scaling of direct Ewald methods at fixed density~\cite{allen2017computer}. In practice, restricting the sum to the top $K$ low-frequency modes yields an $O(N)$ cost for families of comparably sized lattices, with minimal loss of accuracy for many systems. A natural next step is to evaluate the structure factors in a mesh (as in PME/SPME/PPPM) and interpolate the resulting fields to the required $\mathbf{k}$-points.

The combination of FPE with RSA attention provides a promising, fully data-driven route for learning long-range interactions without introducing any ad hoc intermediate quantities (e.g., classical electronegativities or scalar/vector atomic charges). Although accurate, the formulation of reciprocal space attention entails certain limitations. 
Because reciprocal lattice vectors are used during training, the learned representation can exhibit dependence on the chosen $\mathbf{k}$ grid and, by extension, on the lattice boxes encountered during training. In our setting, this dependence appears weak: the model can generalize to moderately large simulation cells and to dense $\mathbf{k}$ grids at inference with minimal loss of accuracy. It is indeed possible to reformulate the RSA in a robust scale-invariant manner by employing a fixed numerical grid and fractional coordinate spaces as done previously by Ewald message passing~\cite{kosmala2023}. We leave this extension for future work.

\section{Conclusion}

We introduced Reciprocal Space Attention (RSA), a k-space attention framework for machine learning interatomic potentials that captures long-range interactions without relying on intermediate observables like charges or empirical corrections. A core component of RSA is Fourier Positional Encoding (FPE), which inherently encodes periodicity and relative atomic positions via Bloch phase factors. FPE when combined with linear attention-like mechanism, arrives at the RSA method which respects translational invariance of the atomic system. Benchmarks on $\text{S}_\text{N}2$ reactions, pair of charged \& polar dimers, random charge systems, liquid NaCl, layered phosphorene and bulk water demonstrate that RSA consistently recovers the long-range asymptotics absent in local and semi-local models. These long-range effects are common in atomistic simulations of molecules and materials, particularly in organic electrolytes, aqueous solutions, and interfacial systems. Our results highlight RSA as a general strategy for integrating long-range interactions into MLIPs such as MACE, thereby extending their applicability to heterogeneous chemical and materials systems.

\section{Methods}
\subsection{Model Implementation}

MPNNs, introduced by Gilmer et al. \cite{gilmer17a}, provide a unifying framework for many GNN architectures for chemical systems. A molecule or crystal is represented as a graph whose nodes are atoms with feature vectors \(\mathbf{h}_m \in \mathbb{R}^{H}\) and positions \(\mathbf{r}_m \in \mathbb{R}^{3}\). Two atoms \(m\) and \(n\) are connected if \(n \in \mathcal{N}(m)\) and \(\|\mathbf{x}_n-\mathbf{x}_m\| < r_{\mathrm{cut}}\), where \(r_{\mathrm{cut}}\) is the distance cutoff. Then MPNN models interactions by iteratively passing and aggregating messages over the edges of the graph. For a given layer $t$, 

\begin{equation*}
\mathbf{M}_m^{(t+1)}
  = \sum_{n\in\mathcal{N}(m)}
    f_{\mathrm{int}}\!\big(\mathbf{h}_m^{(t)},\,\mathbf{h}_n^{(t)},\,\mathbf{e}_{mn}\big),
\end{equation*}
\begin{equation*}
\mathbf{h}_m^{(t+1)}
  = f_{\mathrm{upd}}\!\big(\mathbf{h}_m^{(t)},\,\mathbf{M}_m^{(t+1)}\big).
\end{equation*}
Here, the message sum \(\mathbf{M}_m^{(t+1)}\) aggregates information from the neighborhood
\(\mathcal{N}(m)\) via the interaction function \(f_{\mathrm{int}}\), which depends on the
current node embeddings and edge features \(\mathbf{e}_{mn}\). The learnable update function \(f_{\mathrm{upd}}\) then processes the aggregated messages to produce the next-layer node state. A final readout stage maps the embeddings to task outputs (e.g., atom-wise energy contributions).

In SR MACE, each MP layer consists of a local MP block. In our
LR-augmented MACE, the node features are processed in parallel by a local MP block
and an RSA block. Then their contributions are summed as shown in Fig.~\ref{fig:model_architecture}). For layer \(t\),

\begin{equation*}
\mathbf{M}_{m,\mathrm{nl}}^{(t+1)}
  = \mathrm{RSA}\!\big(\mathbf{H}^{(t)},\,\boldsymbol{\delta}\big)_m,
\end{equation*}
\begin{equation}
\mathbf{h}_{m, nl}^{(t+1)} = f_{\mathrm{upd}}\!\big(\mathbf{h}_m^{(t)},\,\mathbf{M}_{m, nl}^{(t+1)}\big).
\end{equation}

Here, \(\mathbf{H}^{(t)}=(\mathbf{h}_1^{(t)},\ldots,\mathbf{h}_N^{(t)})\) stacks the node
embeddings in the layer \(t\), and \(\boldsymbol{\delta}=(\mathbf{r}_1,\ldots,\mathbf{r}_N)\) represents the set of atomic positions. 

\subsection{Implementation of RSA}
Given atomic features $\mathbf{h}_m \in \mathbb{R}^{H}$, queries, keys, and values are computed as
\begin{align*}
\mathbf{Q}_m &= \mathbf{h}_m \mathbf{W}_Q, \\
\mathbf{K}_m &= \mathbf{h}_m \mathbf{W}_K, \\
\mathbf{V}_m &= \mathbf{h}_m \mathbf{W}_V, 
\end{align*}

where $\mathbf{W}_Q, \mathbf{W}_K, \mathbf{W}_V \in \mathbb{R}^{H \times D}$ are separate weight matrices projecting the vectors to \textit{D} dimension (in our case, \textit{D} is set to \textit{H}). As discussed previously, in this work we consider only the invariant MACE model. So, node features do not have additional dimensions to account for degree $\ell$ and parity $p$.  
Then, according to Eq.~\ref{eq:fpe}, periodic positional representations of the above node vectors can be encoded through FPE.

\begin{align*}
\tilde{\mathbf{Q}}_{m} = \operatorname{FPE}\!\left(\vec r_{m},\, \phi(\mathbf{Q}_{m})\right), \\
\tilde{\mathbf{K}}_{m} = \operatorname{FPE}\!\left(\vec r_{m},\, \phi(\mathbf{K}_{m})\right).
\end{align*}

where $\phi$ denotes a feature map and $\tilde{\mathbf{Q}}, \tilde{\mathbf{K}} \in \mathbb{R}^{k\times H}$, with $k$ the number of vectors in reciprocal space (k space). The present implementation is restricted to orthogonal unit cells. We use SiLU activation~\cite{silu} here,   as the MACE backbone is invariant in this work.

Finally, we evaluate RSA by first calculating the outer product between the keys and values, $\tilde{\mathbf{K}}$ and $\mathbf{V}$ - an $O(N)$ operation computed once per atomic graph in the training set. This is followed by the left multiplication by the query vector $\tilde{\mathbf{Q}}_m$, which evaluates a global attention kernel centered at the reference atom \textit{m} and propagated to all atoms, including their periodic images under periodic boundary conditions. Because each step is linear in system size, the overall cost of RSA remains $O(N)$, as shown in Eq.~\eqref{eq:vm}.

\begin{align}
    \text{RSA}_m(\mathbf{Q, K, V}) 
    &\simeq \sum_{\mathbf{k \neq 0}} 
    \mathbf{w_k} ~\text{FPE}(\phi(\mathbf{Q}_m), \mathbf{r_m})^T  \notag \\
    &\quad \times \sum_{n=1}^N 
    \text{FPE}(\phi(\mathbf{K}_n), \mathbf{r_n}) \, \mathbf{V_n}^T
\end{align}

\subsection{Loss function}
We optimize on the weighted sum of the mean absolute errors (MAE) on energies and forces. Let \(\mathcal{D}\) denote a mini-batch of structures and, for each structure \(\mathcal{S}\in\mathcal{D}\) with \(|\mathcal{S}|\) atoms, let \(\hat{E}^{(\mathcal{S})}\) and \(E^{(\mathcal{S})}\) be the predicted and target energies and \(\hat{\mathbf{F}}^{(\mathcal{S})}_i\) and \(\mathbf{F}^{(\mathcal{S})}_i\) the predicted and target forces for atom \(i\). The loss is

\begin{equation}
\begin{aligned}
\mathcal{L} &=
\frac{1}{|\mathcal{D}|}\sum_{\mathcal{S}\in\mathcal{D}}
\left| \hat{E}^{(\mathcal{S})} - E^{(\mathcal{S})} \right| \\
&\quad + \frac{\lambda}{|\mathcal{D}|}\sum_{\mathcal{S}\in\mathcal{D}}
\frac{1}{|\mathcal{S}|}\sum_{i\in\mathcal{S}}
\left\| \hat{\mathbf{F}}^{(\mathcal{S})}_i - \mathbf{F}^{(\mathcal{S})}_i \right\|_{1}
\end{aligned}
\label{eq:energy-force-loss}
\end{equation}

where \(\lambda\ge 0\) controls the relative weight of the force term. (The \(\ell_1\) norm corresponds to MAE; other norms \(\|\cdot\|_p\) can be used if desired.)

\subsection{MD of Bulk Water}
The simulations of bulk water were performed at a density of 1~g/mL using 300 water molecules in a cubic simulation cell. Trajectories were propagated in the NVT ensemble with a time step of 1~fs, employing Langevin thermostats as implemented in the Atomic Simulation Environment (ASE)~\cite{HjorthLarsen2017}. We ran 10 independent trajectories for both the LR and SR models, each at least 300 ~ps in length, to ensure adequate sampling and statistical convergence.
The generated trajectories were used for subsequent analysis of structural (RDFs) and dielectric properties.

\begin{acknowledgements}
The authors gratefully acknowledge Stephen J. Cox, Ilyes Batatia, and Marcel F. Langer for insightful discussions on model implementation and numerical stability. The authors also thank Stephen J. Cox for his assistance with computing dipole densities and for generously providing the annotated dipole–density analysis code.
This research used resources of the Argonne Leadership Computing Facility, which is a U.S. Department of Energy Office of Science User Facility operated under contract DE-AC02-06CH11357. A.V.M. was supported by the Office of Science, U.S. Department of Energy, under contract DE-AC02-06CH11357.
\end{acknowledgements}

\bibliography{references}
\end{document}